\documentclass{evn2004}
\usepackage{txfonts}
\begin{document}
   \title{Kinematics and dynamics of relativistic jets on large and small
   scales}

   \author{R.A. Laing\thanks{E-mail: rlaing@eso.org}}

   \institute{ESO, Karl-Schwarzschild-Stra\ss e 2, D-85748
   Garching-bei-M\"{u}nchen, Germany}

   \abstract{Modelling of deep VLA images of the jets in FR\,I radio galaxies
has allowed us to derive their three-dimensional distributions of velocity,
emissivity and magnetic-field structure on kiloparsec scales.  By combining our
models of jet kinematics with measurements of the external pressure and density
derived from {\sl Chandra} observations, we can also determine the jet dynamics
via a conservation-law approach. The result is a detailed and quantitative
picture of jet deceleration. We discuss the potential application of these
techniques on VLBI scales. Our fundamental assumption is that the jets are
intrinsically symmetrical, axisymmetric, relativistic flows.  Although this is
likely to be a good approximation on average, the effects of non-stationarity in
the flow (e.g. shocks/knots) may limit its applicability on pc scales. We also
stress the need for both the main and counter-jet to be detected with good
transverse resolution in linear polarization as well as total intensity. Two
foreground effects (free-free absorption and Faraday rotation) must also be
corrected. We comment on the implications of observed VLBI polarization and
Faraday rotation for the field structure (and, by implication, collimation) of
pc-scale jets.  Where our VLA observations start to resolve the jets, about 1
kpc from the nucleus, they are travelling at about 0.8$c$. We briefly discuss the
deceleration of jets from a much faster initial speed.}

   \maketitle
%

\section{Introduction}
\label{Intro}

The determination of physical parameters for the non-thermal plasma in radio
jets is a notoriously difficult problem. This paper describes a new approach
which has met with some success in quantifying the three-dimensional
distributions of velocity, emissivity and field structure of jets in low-power
radio galaxies and outlines potential applications to jets on scales observable
with VLBI. 

\section{Physical principles}
\label{Physics}

\subsection{Outline of the method}

We make the basic assumption that the jets are intrinsically symmetrical and
axisymmetric close to the nucleus, so the observed brightness and polarization
differences between the main and counter-jets are due entirely to the effects of
relativistic aberration.  We then make parameterized models of velocity, field
structure and emissivity, including variation both along and across the jets,
and determine the parameters by fitting to observed Stokes $I$, $Q$ and $U$.  We
calculate the emission by integration along a line of sight on a grid of points,
convolve with the observing beam at one or more resolutions and evaluate
$\chi^2$ over defined areas, using estimates of the small-scale fluctuations
in emission as ``noise levels''. The model parameters can then be optimized
using the downhill simplex method. When evaluating the emissivities in $I$, $Q$
and $U$, we use the techniques described in Laing (\cite{Laing02}), taking proper
account of the effects of relativistic aberration.  Full details are given by
Laing \& Bridle (\cite{LB02a}) and Canvin \& Laing (\cite{CL04}).

\subsection{The importance of linear polarization}

For cylindrical constant-velocity jets emitting isotropically in their rest
frames, the ratio of observed flux density per unit length for the main and
counter-jets depends on both the velocity $\beta c$ and the angle to the line of
sight, $\theta$ via the well-known relation:
\begin{eqnarray*}
 \frac{ I_{\rm j}}{I_{\rm cj}} & = & \left (\frac{1 + \beta\cos\theta}{1 -
  \beta\cos\theta} \right )^{2+\alpha} \\
\label{eq:iratio}
\end{eqnarray*}
where $\alpha$ is the spectral index.  In order to break the degeneracy between
$\beta$ and $\theta$, we use the linear polarization.  The relation between the
angles to the line of sight in the rest frame of the flow, $\theta^\prime$ and
in the observed frame, $\theta$, is:
\begin{eqnarray*} 
\sin\theta^\prime_{\rm j} & = & [\Gamma(1-\beta\cos\theta)]^{-1}\sin\theta
\makebox{~~~~~(main jet)} \\
\sin\theta^\prime_{\rm cj} & = & [\Gamma(1+\beta\cos\theta)]^{-1}\sin\theta
\makebox{~~(counter-jet)} \\
\end{eqnarray*}
($\Gamma$ is the Lorentz factor). The observed polarization is in general a
function of $\theta^\prime$.  The observed distributions of the main and
counter-jets can therefore be very different (see Section~\ref{FRI} for an
example).  If we know the field structure a priori, then we can solve explicitly
for $\beta$ and $\theta$, but in general, we must fit the field configuration
and therefore need to introduce two additional parameters to describe its
anisotropy. These also determine the variation of polarization transverse to the
jet axis, and can be estimated independently of the velocity and angle if the
jets are well resolved in this direction.

\subsection{Ordered and disordered fields}
\label{Border}

Our modelling can determine the ratio of the components of magnetic field
(defined as rms values along given directions) but can only partially
constrain whether they are vector-ordered or have many small-scale reversals. We
assume that the fields have zero mean with many reversals on scales smaller than
the observing beam, but are made anisotropic by shear or compression (Laing
\cite{Laing80,Laing02}). If {\em one} of the three field components is
vector-ordered, however, our results are unchanged. For example, a jet with a
confining (ordered) toroidal field but disordered radial and longitudinal
components could not be distinguished by its synchrotron emission from one in
which all three components are disordered. [Internal Faraday rotation can be
used to differentiate between these cases if the internal density of thermal
plasma is high enough]. The situation for a helical field is different (Laing
\cite{Laing81}): unless the jets are observed at 90$^\circ$ to the line of sight
in the rest-frame of the emitting material, their transverse brightness and
polarization distributions are not symmetrical. The very high degree of symmetry
observed in FR\,I jets argues that no more than one of the field components is 
vector-ordered and flux conservation indeed suggests that the longitudinal
component must have many reversals (Begelman, Blandford \& Rees \cite{BBR}).

   \begin{figure}
   \centering
   \vspace{225pt}
   \includegraphics{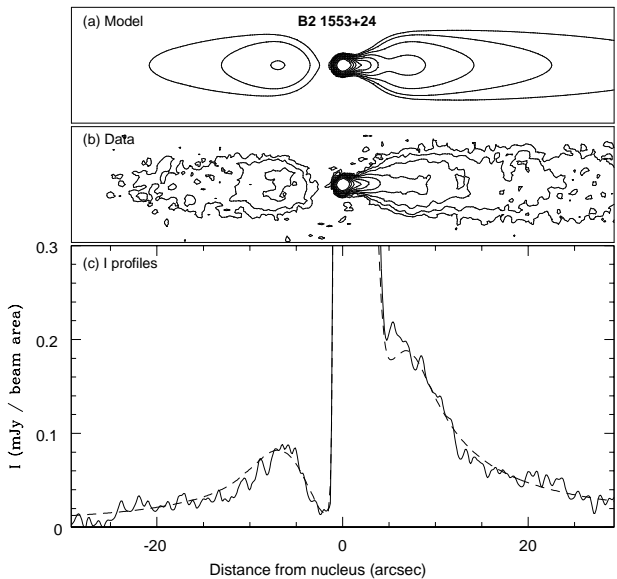}
      \caption{Comparison of model and observed total intensity for
      B2\,1553+24. (a) observed contours; (b) model contours; (c) longitudinal
      profiles for observations (full line) and model (dashed). See Canvin \&
      Laing \cite{CL04} for details.
         \label{fig:1553i}
         }
   \end{figure}

   \begin{figure*}
   \centering
   \vspace{375pt}
   \includegraphics{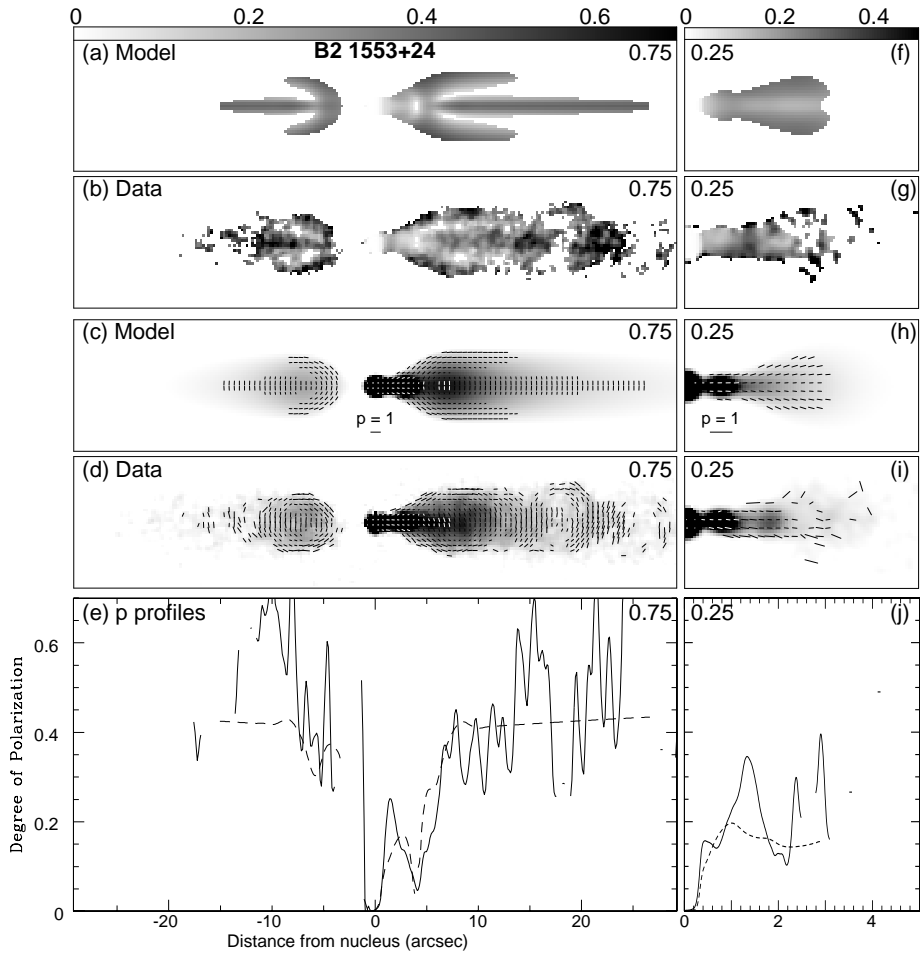}
   \caption{Comparison of model and observed linear polarization for
   B2\,1553+24. (a) -- (e) 0.75\,arcsec FWHM. (a) model and (b) observed degree
   of polarization. (c) model and (d) observed vectors showing the degree of
   polarization and the apparent magnetic field direction, superposed on
   grey-scales of $I$. (e) longitudinal profile of degree of
   polarization. Panels (f) -- (j) show the same quantities at 0.25\,arcsec
   resolution. Note the large differences in polarization between the main and
   counter-jets (panels a - d); these are successfully modelled as an effect of
   relativistic aberration.
            \label{fig:1553lp}
           }
    \end{figure*}
\section{Application to FR\,I jets on kiloparsec scales}
\label{FRI}

\begin{figure}
   \centering
   \vspace{400pt}
   \includegraphics{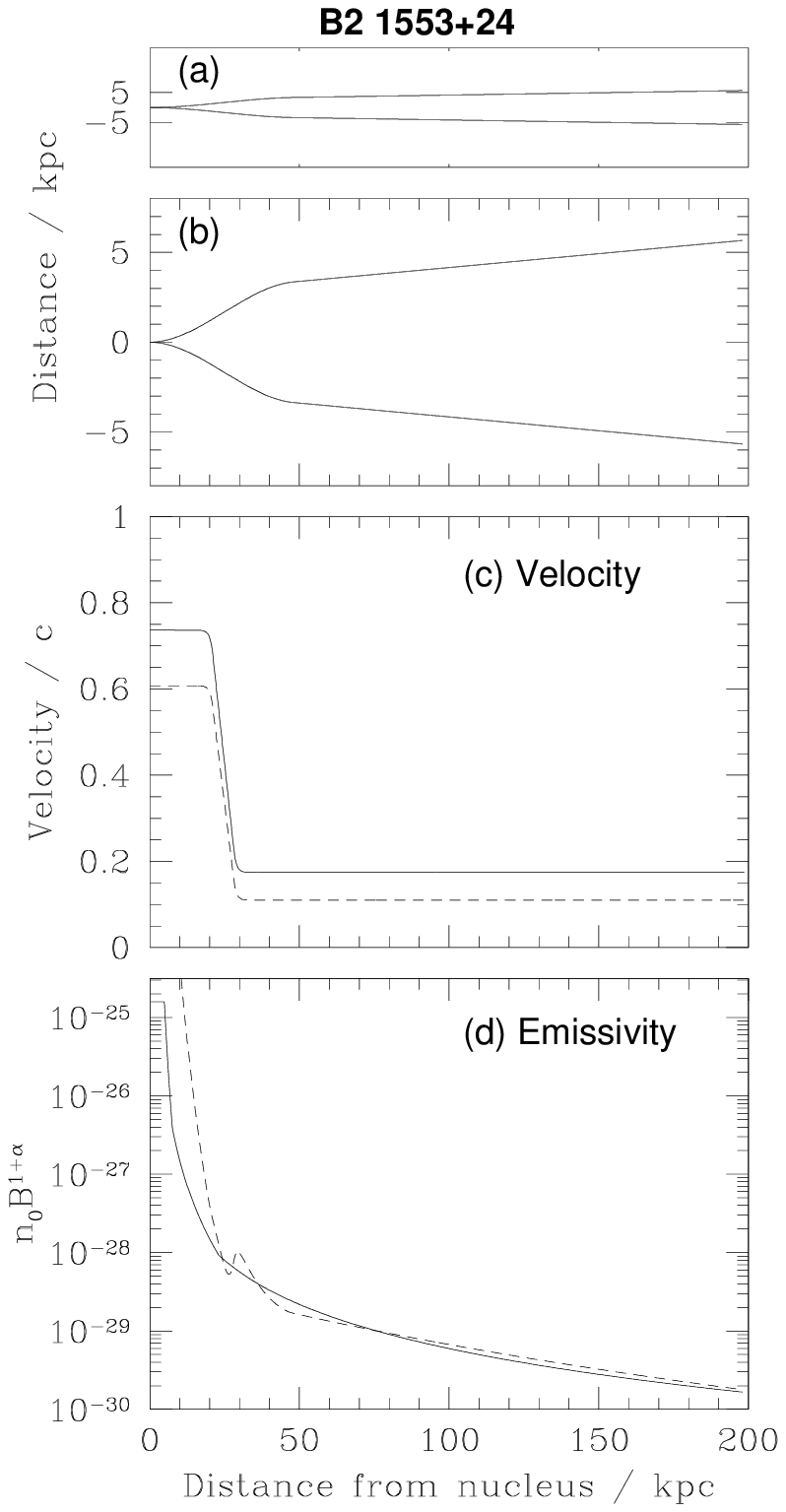}
      \caption{Inferred variation of physical quantities along the jets of
      B2\,1553+24. (a) and (b) jet geometry (note the different scales); (c)
      on-axis (full) and edge (dashed) velocity; (d) proper emissivity (full
      line; the dashed line shows a quasi-one-dimensional adiabatic model).
         \label{fig:1553prof}
         }
   \end{figure}

\begin{figure}
   \centering
   \vspace{400pt}
   \includegraphics{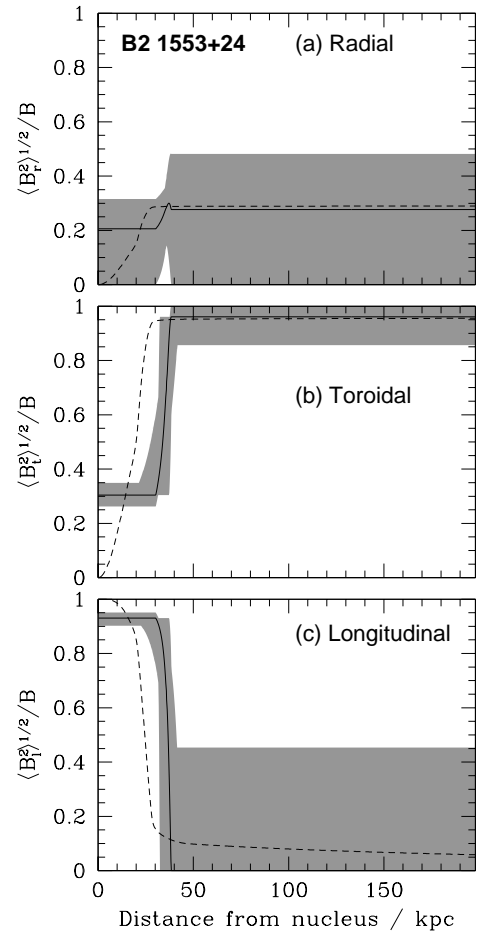}
      \caption{Profiles of the relative strengths of the three orthogonal
      magnetic-field components in B2\,1553+24. The full lines show the best
      values and the shaded areas the uncertainties. the dashed lines are
      predicted from flux-freezing in a laminar flow. (a) radial; (b) toroidal;
      (c) longitudinal. 
         \label{fig:1553bprof}
         }
   \end{figure}

We have so far completed model-fitting for four sources: 3C\,31 (Laing \& Bridle
\cite{LB02a}), 0326+39 and 1553+24 (Canvin \& Laing \cite{CL04}) and NGC\,315 (Cotton et al.,
in preparation). A recent status report for the project was given by Laing,
Canvin \& Bridle (\cite{LCB}).  
Our principal results to date are as follows.
\begin{enumerate}
\item An intrinsically symmetrical, relativistic jet model provides an excellent
description of the total intensity and linear polarization observed in the four
sources studied so far (e.g.\ Figs~\ref{fig:1553i} and \ref{fig:1553lp}).
\item We can estimate the angle to the line of sight and the three-dimensional
distributions of velocity, emissivity and magnetic-field structure. The jets in
all of the sources decelerate from $\beta \approx 0.8$ to $\beta \approx$ 0.1 --
0.2 over short distances within the region of rapid expansion. Further out they
recollimate and subsequent deceleration is slow or absent. The ratio of edge to
on-axis velocity is consistent with 0.7 everywhere [Fig.~\ref{fig:1553prof}(c)]. 
\item The magnetic field evolves from predominantly longitudinal close to the
nucleus to mainly toroidal at large distances; the behaviour of the (weaker)
radial component differs between the sources (e.g.\ Fig.~\ref{fig:1553bprof}).
\item By combining the kinematic model for 3C\,31 with a description of the
external gas density and pressure derived from {\em Chandra} observations
(Hardcastle et al. \cite{Hard02}) and using conservation of particles, energy and
momentum, Laing \& Bridle (\cite{LB02b}) demonstrated that the jet deceleration could
be produced by entrainment of thermal matter and derived the spatial variations
of pressure, density and entrainment rate for the first time.
\item Laing \& Bridle (\cite{LB04}) applied an adiabatic jet model (including the
effects of shear and arbitrary initial conditions for the magnetic field)
and showed that it provided a fair description of brightness and
polarization at projected distances of 3 -- 10\,kpc from the nucleus, but
failed completely closer in. They showed that additional particle
acceleration is required precisely where the X-ray emission from the jet
is bright. 
\end{enumerate}

\section{Application to parsec-scale jets}
\label{VLBI}

\subsection{Requirements}

In order to model jets using the technique described above, there are several
obvious requirements:
\begin{enumerate}
\item Both the main and counter-jets must be detected and well resolved 
  along and across their lengths. This probably rules out objects at a very
  small angle to the line of sight.
\item Linear polarization must be measurable in both jets and any effects of
  (internal or external) Faraday rotation must be removed.
\item The jets must be close enough to intrinsic symmetry that observed
  differences between them are dominated by the effects of
  aberration.
\item Sources close to the plane of the sky cannot be modelled, as there should
  then be no significant differences between the two jets.
\end{enumerate}
These requirements immediately rule out the majority of the bright core-jet
sources observed with VLBI, but this is hardly a surprise: selection of such
objects is heavily biased by Doppler boosting. Not only are their counter-jets
normally invisible, but the main jets have any bends amplified by projection.
The most suitable targets are sources at less extreme orientations with known
counter-jets, such as Cygnus\,A (Krichbaum et al.\ \cite{CygA1}; Bach et al.\
\cite{CygA2}) and Centaurus\,A (Tingay et al.\ \cite{CenA1}).

\subsection{Complications}

We assume that the jets are intrinsically symmetrical, stationary flows.  It is
a reasonable hypothesis that parsec-scale jets are symmetrical on
average and that environmental asymmetries are no more important than on larger
scales. The flow in parsec-scale jets is certainly not stationary, however. This
may not matter provided that the observations average over significant numbers
of discrete features: all that is necessary is that the properties of the main
and counter-jets are on average the same. If the brightness distributions are
dominated by a few stochastic features (shocks, for example), then the approach
will fail.  It is also essential to remove the effects of free-free absorption
(e.g.\ Tingay \& Murphy \cite{CenA2}) before comparing jet and counter-jet
brightnesses.

We can rule out ordered helical fields for the FR\,I jets we have modelled
(Section~\ref{Border}) but the possibility needs to be re-evaluated for pc-scale
jets (e.g.\ Lyutikov et al.\ \cite{LPG}).  Faraday rotation offers both a threat
and an opportunity. Detection of internal rotation may provide evidence for the
much-sought-after toroidal confining field (Asada et al.\ \cite{Asada}; Gabuzda,
Murray \& Cronin \cite{GMC}) but foreground material, perhaps interacting with
the jet, will confuse the issue (Zavala \& Taylor \cite{ZT}). In
either case, our models depend on accurate measurement of the polarization
uncorrupted by Faraday effects. As with the free-free absorption problem, this
may require observations at very high frequencies.

Where we first determine the velocities of kpc-scale jets in FR\,I sources, they
are $\approx$0.8$c$. Given the strong evidence for faster flow on smaller scales
in these objects (e.g.\ Giovannini et al. \cite{Giov}), the jets must decelerate
significantly on scales of 10\,pc -- 1\,kpc. This region may be accessible to
observation using the EVN and e-MERLIN.   

\section{Conclusions}
\label{Concl}

We conclude that the techniques developed to model FR\,I jets on kiloparsec
scales as intrinsically symmetrical, relativistic flows can in principle be
applied to VLBI observations. The requirements are stringent: both jets must be
imaged in detail in total intensity and linear polarization and propagation
effects such as Faraday rotation and free-free absorption must be corrected. If
such observations are feasible on sources such as Cyg\,A and Cen\,A, they will
allow the three-dimensional distributions of velocity, emissivity and magnetic
field to be determined for the first time.

\begin{acknowledgements}
 The National Radio Astronomy Observatory is a facility
of the National Science Foundation operated under cooperative agreement by
Associated Universities, Inc.
\end{acknowledgements}


\begin{thebibliography}{}

\bibitem[2002]{Asada} Asada, K.\ et al.\ 2002, PASJ, 54, L39

\bibitem[2002]{CygA2} Bach, U., Krichbaum, T.P., Alef, W., Witzel, A. \& Zensus,
  J.A.\ 2002, 6th European VLBI Network Symposium, eds Ros, E., Porcas, R.W.,
  Lobanov, A.P. \& Zensus, J.A., MPIfR, Bonn, 155

\bibitem[1984]{BBR} Begelman, M.C., Blandford, R.D. \& Rees, M.J.\ 1984,
Rev. Mod. Phys., 56, 255

\bibitem[2004]{CL04} Canvin, J.R. \& Laing, R.A.\ 2004, MNRAS,
  350, 1342

\bibitem[1998]{CygA1} Krichbaum, T.P.\ et al.\ 1998, A\&A 329, 873

\bibitem[2004]{GMC} Gabuzda, D.C., Murray, \'{E}.\ \& Cronin, P.\ 2004, MNRAS,
  351, L89

\bibitem[2001]{Giov} Giovannini, G., Cotton, W.D., Feretti, L., Lara, L. \&
  Venturi, T.\ 2001, ApJ, 552, 508
  
\bibitem[2002]{Hard02} Hardcastle, M.J., Worrall, D.M.,
Birkinshaw, M., Laing, R.A. \& Bridle, A.H.\ 2002, MNRAS, 334, 182

\bibitem[1980]{Laing80} Laing R.A.\ 1980, MNRAS, 193, 439

\bibitem[1981]{Laing81} Laing R.A.\ 1981, ApJ, 248, 87 

\bibitem[2002]{Laing02} Laing R.A.\ 2002, MNRAS, 329, 417

\bibitem[2002a]{LB02a} Laing, R.A. \& Bridle, A.H.\ 2002a, 
MNRAS, 336, 328

\bibitem[2002b]{LB02b} Laing, R.A. \& Bridle, A.H.\ 2002b, 
MNRAS, 336, 1161

\bibitem[2004]{LB04} Laing, R.A. \& Bridle, A.H.\ 2004, 
MNRAS, 348, 1459

\bibitem[2004]{LCB} Laing, R.A., Canvin, J.R. \& Bridle, A.H., 2004, New AR, 47,
  435

\bibitem[2004]{LPG} Lyutikov, M., Pariev, V.I. \& Gabuzda, D.C., MNRAS,
  submitted, {\tt astro-ph/0406144}

\bibitem[1998]{CenA1} Tingay, S.J.\ et al.\ 1998, AJ, 115, 960 

\bibitem[2001]{CenA2} Tingay, S.J. \& Murphy, D.W.\ 2001, ApJ, 246, 210

\bibitem[2003]{ZT} Zavala, R.T. \& Taylor, G.B., 2003, ApJ, 589, 126
\end{thebibliography}
\end{document}